\documentclass[12pt,preprint]{aastex}
\usepackage{graphics,epsf}
\usepackage{amsmath}                
\usepackage{amsfonts}               
\usepackage{amssymb}                
\usepackage{epsfig}

\def \eV{~\rm{eV}}

\def \cm{~\rm{cm}}
\def \s{~\rm{s}}
\def \km{~\rm{km}}

\def \K{~\rm{K}}
\def \g{~\rm{g}}

\def \AU{~\rm{AU}}

\def \yrs{~\rm{yrs}}
\def \yr{~\rm{yr}}

\def \kpc{~\rm{kpc}}

\def \days{~\rm{days}}
\def \Jy{~\rm{Jy}}
\def \mum{~\rm{\mu m}}

\begin{document}

\title{ACCRETION ONTO THE COMPANION OF ETA CARINAE DURING THE SPECTROSCOPIC EVENT. V.
\newline
THE INFRARED DECLINE}

\author{Amit Kashi\altaffilmark{1} and Noam Soker\altaffilmark{1}}

\altaffiltext{1}{Department of Physics, Technion$-$Israel
Institute of Technology, Haifa 32000 Israel;
kashia@physics.technion.ac.il;
soker@physics.technion.ac.il.}

\begin{abstract}
We propose that the decline in the near-IR flux from the massive binary
system $\eta$ Carinae during the spectroscopic event might be explained
by accreted mass that absorbs the radiation from the secondary star,
and by that reduces the heating of the dust that is responsible for the
near-IR emission. This binary system has an orbital period of
$2024$~days and eccentricity of $e \simeq 0.9$. The emission in several
bands declines for several weeks near every periastron passages, in
what is termed the spectroscopic event. In the \emph{accretion model}
for the spectroscopic event the secondary star accretes mass from the
primary's wind for $\sim 10~$weeks near every periastron passage. The
mass is accreted mainly in the equatorial plane. The disk and its wind
block the secondary's radiation from heating dust that does not reside
within narrow cones along the symmetry axis. This, we propose, might
explain the decline in the near-IR flux occurring at the beginning of
each spectroscopic event. We also argue that the increase in the
near-IR prior to the event might be accounted for by enhanced hot ($T
\sim 1700 \K$) dust formation in the collision region of the winds from
the two stars. This dust resides within $\sim 60^\circ$ from the
equatorial plane, and most of it cannot be heated by the secondary
during the accretion phase.
\end{abstract}


\section{INTRODUCTION}
\label{sec:intro}

The luminous blue variable (LBV) Eta Carinae has been observed and
widely studied for almost 170 years, but the existence of its binary
companion (Damineli 1996) wasn't fully accepted until about ten years
ago. The two winds blown by the massive stellar binary system are major
players in the $5.54 \yrs$ periodicity, as was recognized in the past
particularly for the X-ray behavior (Corcoran 2005; Pittard \& Corcoran
2002; Akashi et al. 2006). The importance of the secondary star and the
colliding winds in explaining the behavior at other wavelengths has
emerged only recently (Abraham et al. 2005; Ishibashi et al. 1999;
Damineli et al. 2000; Corcoran et al. 2001; 2004; Hillier et al. 2001;
Pittard \& Corcoran 2002; Smith et al. 2004; Steiner \& Damineli 2004;
Verner et al. 2005; van Genderen et al. 2006; van Genderen \& Sterken
2007; Soker 2007; Kashi \& Soker 2007a, b; Soker \& Behar 2006; Behar
et al. 2007).

LBVs are massive stars that are believed to be in a rapid and unstable
evolutionary phase in which tens of solar mass of material are ejected
into the interstellar medium over a relatively short period of time,
possibly in eruptive events (Smith \& Owocki 2006; Smith 2007). Such a
twenty year long outburst, known as the `Great Eruption', was
experienced by $\eta$ Car in 1843-1863 (Davidson \& Humphreys 1997),
accompanied by ejection of a mass of $>12 M_\odot$ (Smith et al. 2003a;
Smith \& Owocki 2006; Smith 2007). This ejecta now form a reflecting
bipolar nebula$-$the `Homunculus'$-$which has a dense equatorial gas,
and dense clumps above and below the equatorial plane that obscure the
central star at many wavelengths. Another outburst, the `Lesser
Eruption' occurred between 1887 and 1895, and ejected material mainly
in the polar direction, but also in the equatorial direction (Smith
2005; Smith 2005 \& Gehrz 1998; Humphreys et al. 1999).

The ejected mass has cooled since the Great Eruption, and formed $\sim
0.1 M_\odot$ of dust, which obscures the central star, reflects light
toward us in other directions, and emits in the IR band (Smith et al.
2003b). The IR luminosity is $L_{IR} \simeq 5 \times10^6 L_{\odot}$
(Cox et al. 1995; Smith et al. 2003b). According to estimates made by
Davidson \& Humphreys (1997), this IR emission comprises $\sim 90 \%$
of the luminosity emitted by $\eta$ Car.

The light curve of $\eta$ Car  is characterized by a roughly
constant luminosity at all bands, although with fluctuations, for
most of the $5.54 \yrs$ orbital period (Whitelock et al. 2004;
Falceta-Goncalves et al. 2005; Abraham et al. 2005; Corcoran 2005;
Duncan \& White 2003; Nielsen et al. 2007). Large variations,
typically with slow increase and then rapid decrease that follows by
a low state that typically lasts tens of weeks, occur every
$2024$~days in what is termed the spectroscopic event, after the
disappearance of highly ionized lines (Damineli 1996; Damineli et
al. 2000). The event is thought to occur near periastron passage,
where the distance between the two stars is $r=1.66\AU$ (for the binary parameters
used by us).
In the last periastron passage in 2003.5, for example, the sharp X-ray drop
occurred in $\rm{JD} 2452819.6$ (Corcoran 2005).

In this paper we address the behavior of the near-IR light-curve
before and during the spectroscopic event, attributing the sharp
decline in the near-IR to an accretion event. This is the $5^{th}$
paper in a series of papers aiming at understanding some of the
processes during the spectroscopic event by a $\sim 10~$weeks long
accretion event onto the secondary star. The accretion event was
proposed to explain the X-ray behavior (Soker 2005; Akashi et al.
2006; Soker \& Behar 2006), and was applied to the visible band as
well (Soker \& Behar 2006; Kashi \& Soker 2007b; Soker 2007).

\section{THE INFRARED BEHAVIOR NEAR PERIASTRON PASSAGE}
\label{sec:models}

\subsection{The light curve in the near-IR}
\label{sec:light_curve}

\begin{figure}[h!]
\resizebox{0.99\textwidth}{!}{\includegraphics{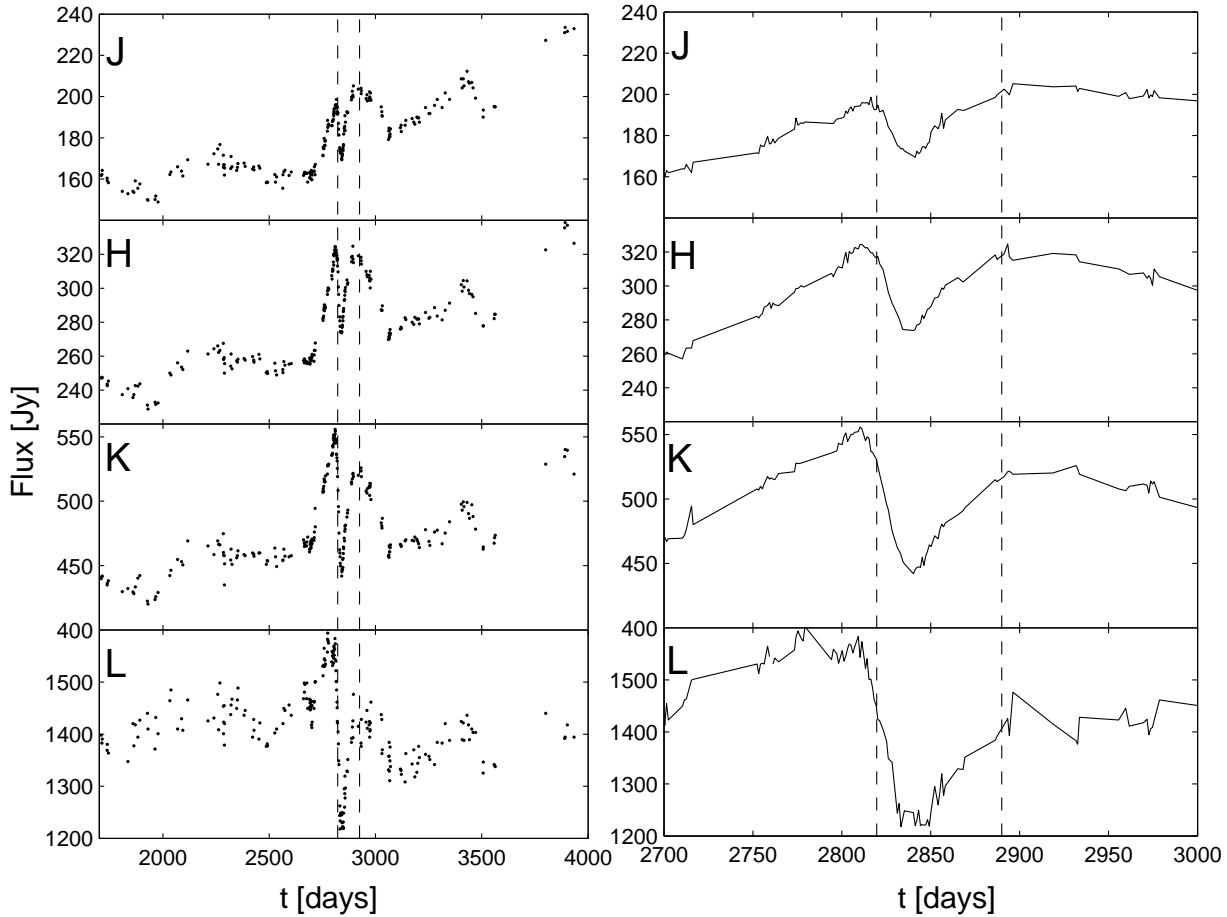}}
\caption{Left: The \emph{JHKL} fluxes for the last cycle taken from
Whitelock et al. (2004; Whitelock, P. 2007 private communication), from
top to bottom, respectively. Right: Zooming on the $2003.5$ event. The
left vertical lines in both sides of the figure mark the X-ray decline
(Corcoran 2005). The right vertical lines mark the end of the assumed
10 week-long accretion process (Soker 2005; Akashi et al. 2006) as
discussed in section \ref{sec:accretion}.} \label{whitelock}
\end{figure}

In this section we will describe several possible explanations for the
near-IR behavior, which we however, find to be problematic. We start by
describing the behavior of the near-IR in the \emph{JHKL} bands, from
Whitelock et al. (2004; Whitelock, P. 2007 private communication) who
presented a detailed near-IR survey of the \emph{JHKL} bands, which was
taken for a period of almost 40 years. The flux of the last cycle (the
2003.5 event) can be seen in Fig. \ref{whitelock}. The main
characteristics of the last cycle are as follows:
\begin{enumerate}
\item \textbf{Typical flux values.} The longer the wavelength the larger are
the average quiescent flux (the flux during most of the time when
the system is away from periastron) and the variations in the flux.
The average quiescent flux in \emph{J} is $\sim 160 \Jy$, while that
in \emph{L} is $\sim 1440 \Jy$.
\item \textbf{Rapid increase.} Approximately $\sim 150$~days before the event
a rapid increase in the flux in all wavelengths begins. In \emph{J} and
\emph{L} the fluxes reach peak values which are $\Delta F_J \simeq 40
\Jy$ and $\Delta F_L \simeq 170 \Jy$ above the average quiescent flux.
\item \textbf{The event.} During the event the hard X-ray emission
drops to $\sim 1\%$ of the flux before minimum and stays close to that
level for about 70 days (Corcoran 2005; Hamaguchi et al. 2007). The IR
event is preceded by a brightening at all near-IR wavelengths, as is
the case for the X-ray light curve. However, in marked contrast to the
X-ray light curve the near-IR fluxes decrease during the event by only
$\sim 15-24 \%$. Only in the \emph{L} band the light curve is
flat-bottom, and only for $\sim 20$~days. While the oscillations in the
X-ray flux make it difficult to pinpoint the time of ingress, it is
clearly earlier than it is at IR wavelengths. Within the near-IR light
curves ingress starts earliest at \emph{L} and latest at \emph{J} where
it is more or less coincident with the well-determined start of X-ray
flat minimum; the delay between \emph{L} and \emph{J} is $\sim 7$~days.
The \emph{K} minimum occurs at X-ray phase 0.011 ($\sim 22$~days after
the beginning of the flat X-ray minimum). We will attribute the delay
of the \emph{J} ingress to the formation of extra hot dust close to the
secondary star near periastron passages.
\item \textbf{Typical variation amplitude.}
Both the relative and absolute decline (the V-shaped part of the light
curves) during the event are larger for longer wavelengths, being
$\Delta F_J \simeq -35 \Jy$, and $\Delta F_L \simeq -400 \Jy$ from peak
to minimum. In the \emph{L} band the minimum flux in the event is below
the average quiescent flux.
\item \textbf{Recover.}
After the sharp decline, the \emph{L} band recovers to more or less
its average quiescent flux, while in the \emph{JHK} bands the fluxes
recover to values above the average quiescent flux, and only after
the event (lasting $\sim 10$~weeks) they decline slowly to the
quiescent level.
\end{enumerate}

\subsection{Problems with a free-free emission model}

\begin{figure}[h!]
\resizebox{0.49\textwidth}{!}{\includegraphics{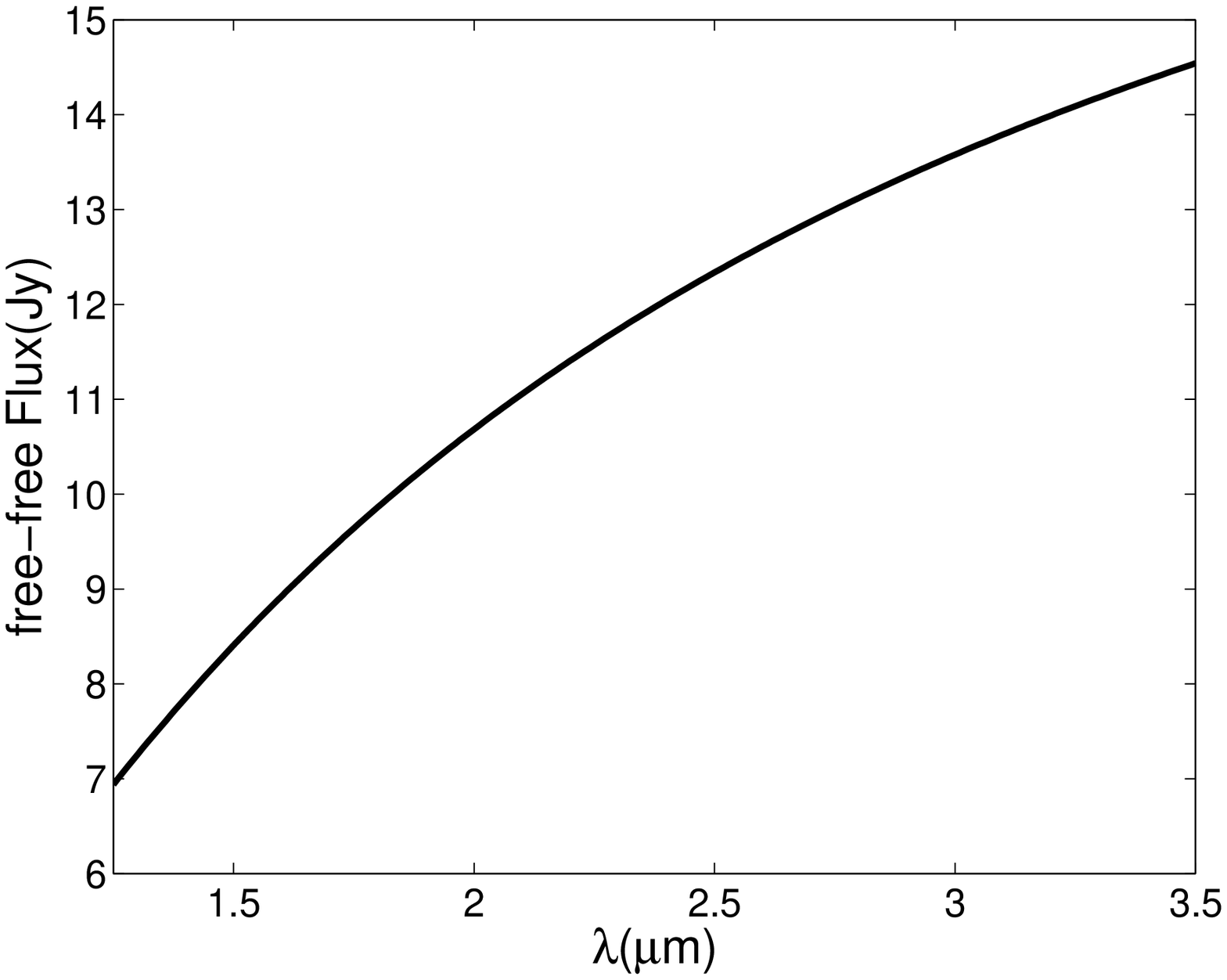}}
\caption{The free-free flux in the near-IR as given by equation
\ref{F1}.} \label{ffFlux}
\end{figure}

Whitelock et al. (2004) argued that the main contribution to the
\emph{JHKL} flux variations are due to free-free emission. We now show
that this explanation is problematic since free-free emission cannot
provide enough flux to explain the observations. We assume that the
recombination rate $\dot{R}=\alpha_B \int n_e n_i dV$, is in
equilibrium with the ionizing photon emitted by the secondary star per
unit time $\dot \Phi_{2}=2.5-4.4\times10^{49} s^{-1}$ (see Kashi \&
Soker 2007a). Here $n_e$, $n_i$, and $dV$ are the electron number
density, ions number density and a unit volume, respectively. Using
this assumption to substitute for $\int n_e n_i dV$ in the expression
for the free-free emission in the near-IR (Rybicki \& Lightmann 2004)
we find the contribution of free-free emission to the observed flux
\begin{equation}
F=22 \left(\frac{\dot\Phi_{2}}{4.4\times10^{49}
\rm{s^{-1}}}\right) \exp\left[{-1.44\left(\frac{\lambda}{1\mu
m}\right)^{-1}}\right]{\rm Jy}. \label{F1}
\end{equation}
We took $T=10^4 \K$ in equation (\ref{F1}) (for other values of T the
emission will be about the same or lower), and $D_\eta=2.3 \kpc$ as the
distance to $\eta$ Car (Davidson \& Humphreys 1997). We also used
$g_{ff}\simeq1.2$ as the Gaunt factor in the near-IR. The expected flux
in the near-IR as given by equation (\ref{F1}) is plotted in Fig.
\ref{ffFlux}.

Fig. \ref{ffFlux} shows that the fluxes the free-free emission can
supply, e.g., $F_J=5.8 \Jy$ and $F_L=12 \Jy$, are much smaller than the
typical variations in fluxes as can be seen in Fig. \ref{whitelock}.
The ionization by the primary star was not taken into account. Adding
it to the ionizing photon rate of the secondary star would increase the
flux by $20\%$ at most. Moreover, these results were obtained using
$100\%$ production efficiency, and the maximum accepted value for the
recombination rate $\dot{R}$, so we expect the free-free contribution
to the near-IR flux to be even smaller. We conclude that the free-free
emission cannot explain the near-IR behavior.

We emphasize that our calculated changes in the flux near periastron
are attributed to ionization by the secondary. The changes, according
to the free-free model we are examining here, are due to the dense gas
in the colliding wind region. This gas must be ionized. The primary is
much cooler than the secondary, and despite being five times as
luminous as the secondary has a much lower ionizing flux. Fig.
\ref{ffFlux} shows that the ionizing radiation cannot explain the
changes. More problematic to the free-free emission model is the fact
that both the quiescent flux and the variations are larger in the
\emph{L} band than in the \emph{K} band. The free-free emission from
the primary wind is expected to behave in the opposite manner as in the
case of the wind of P Cygni (Lamers et al. 1996). Therefore, although
the primary wind does contribute to the near-IR emission via free-free
emission, it cannot be the main near-IR source. Indeed, in P Cygni the
near-IR flux is two orders of magnitude below that in Eta Car (Lamers
et al. 1996).

\subsection{Problems with a free-free absorption}

We now show that the free-free absorption cannot lead to sharp decline
in the near-IR during the event. The absorption can be explained in
principle by the dense conically-shaped post-shock primary wind. The
decline during the event in \emph{J} is $\sim 15 \%$ and in \emph{L} it
is $\sim 24 \%$ (section \ref{sec:light_curve}). The free-free optical
depth in the \emph{L} band is $\sim 8.4$ times larger than that in
\emph{J}. Therefore, extinction by $\sim 15 \%$ in the \emph{J} band
implies almost completely extinction of the \emph{L} band. The similar
decline in \emph{J} and \emph{L} can be solved in principle if the
dense material covers all (or most) of the hot dust emitting in the
\emph{J} band, and absorbs $\sim 15 \%$ of it. The same dense gas
covers and completely absorbs only $\sim 24\%$ of the radiation from
the warm dust responsible for the \emph{L} band. This requires a
special geometry.

Another problem with such an absorbing model is that there is no
good candidate for an absorber:

\begin{enumerate}
\item The dense primary wind close to the primary covers a too small
angle, considering that the orbital plane is inclined by $\sim 45
^\circ$ to the line of sight, and the IR emitting region is expected
to be large.

\item The dense post-shock primary wind
material (the conical shell) cannot absorb enough of the
\emph{J}-band emission. For details see Appendix A.

\item The ejection of a dense shell as a general explanation for
the spectroscopic event of $\eta$ Car (Zanella et al. 1984) is ruled
out by the X-ray behavior (Akashi et al. 2006).

We therefore regard any model for the sharp decline in the near-IR that
is based on a dense gas that obscures the IR emitting regions as
unlikely.

\subsection{Problems with decreasing dust formation rate}

Another model that can account in principle for the sharp decline in
the near-IR is based on a fast decline in dust formation during the
event. Although we do argue for near-IR emission from dust, an
explanation based on a decrease in the dust formation rate for the
near-IR decline cannot work for $\eta$ Car. The decline in the dust
formation rate occurs because of a decline in the supply of matter from
the binary system and/or an enhanced emission that prevents the matter
from cooling and forming dust. In either cases we would expect that the
dust residing closer to the binary system will be influenced first,
namely, the hotter dust. The hotter dust contributes to the shorter IR
wavelengths, and would imply that the \emph{J}-band decline occurs
before the \emph{L}-band decline. This is contrary to observation
(Whitelock et al. 2004; see Fig. \ref{whitelock} here).

\end{enumerate}

\section{FITTING THE LIGHTCURVE}
\label{sec:fit_lightcurve}

Cox et al. (1995) explained the IR light curve of $\eta$ Car by a
modified blackbody emission from dust at two temperatures: A cold dust
component at $210 \K$, and a warm dust component at $430 \K$. We are
interested in the hot dust, and for that we will not modify the fitting
of the cold dust as was done by Smith et al. (2003b; see below), but
rather only add a third hot component, to better fit the \emph{JHKL}
bands. We will use the same two dust components and the modified
blackbody function as in Cox et al. (1995)
\begin{equation}
\tilde{B}(T)= Q B(T) = \left(\frac{\lambda}{\lambda_0}\right)^{-m} B(T),
\label{tildeB}
\end{equation}
where $B(T)$ is the usual Planck function, and $m=1$ for amorphous
Carbon grains (Hildenbrand 1983).

In choosing the temperature of the hot dust component we note that
Smith et al. (2007) used hot dust at $T=1660-1750 \K$ to explain the
emission from SN2006jc. They claimed that although the shape of the red
continuum excess can be approximately fit by a $T \sim 2000 \K$ Planck
function, optically thin emitting dust grains will have
wavelength-dependent emissivity and as long as the grains are not
larger than about $a=1\mum$, the dust will typically have emissivity
proportional to $\lambda^{\beta}$ with $-1\leq\beta\leq-2$. This model
is similar to the modified blackbody function of Cox et al. (1995). The
work of Smith et al. (2007) suggests that dust can also reach
temperatures of  $T \sim 1700 \K$ in an environment similar to that in
the colliding winds of $\eta$ Car. Following the above discussion,  we
will take the dust in $\eta$ Car to be composed of three components:
Hot dust at $T=1700 \K$, warm dust at $T=430 \K$, and cold dust at $T=
210 \K$. We emphasize that we expect the dust in $\eta$ Car to have a
continuous range of temperatures, and therefore the three dust
components used by us should be regarded as a representative
description of the real situation (a `toy model'). Smith et al.
(2003b), for example, fit the cold dust with two temperatures of $200
\K$ and $140 \K$. We are not interested in the cold dust in this work,
and present the cold dust with only one temperature. For that we chose
a temperature of $210 \K$. Our fit cannot be used for a quantitative
study of the cold gas. In addition, different points in the lightcurve
are taken from different sources (as summarized in Cox et al. 1995).
For these reasons our treatment of the cold dust is not accurate, and
the work of Smith et al. (2003b) that is superior to our fit should be
consulted when studying the cold dust. This approximate description is
sufficient to teach us about the basic physical processes relevant to
the near-IR emission. Only the two hotter components, the hot and the
warm, are relevant to the \emph{JHKL} wavebands.

\begin{figure}[h!]
\resizebox{0.89\textwidth}{!}{\includegraphics{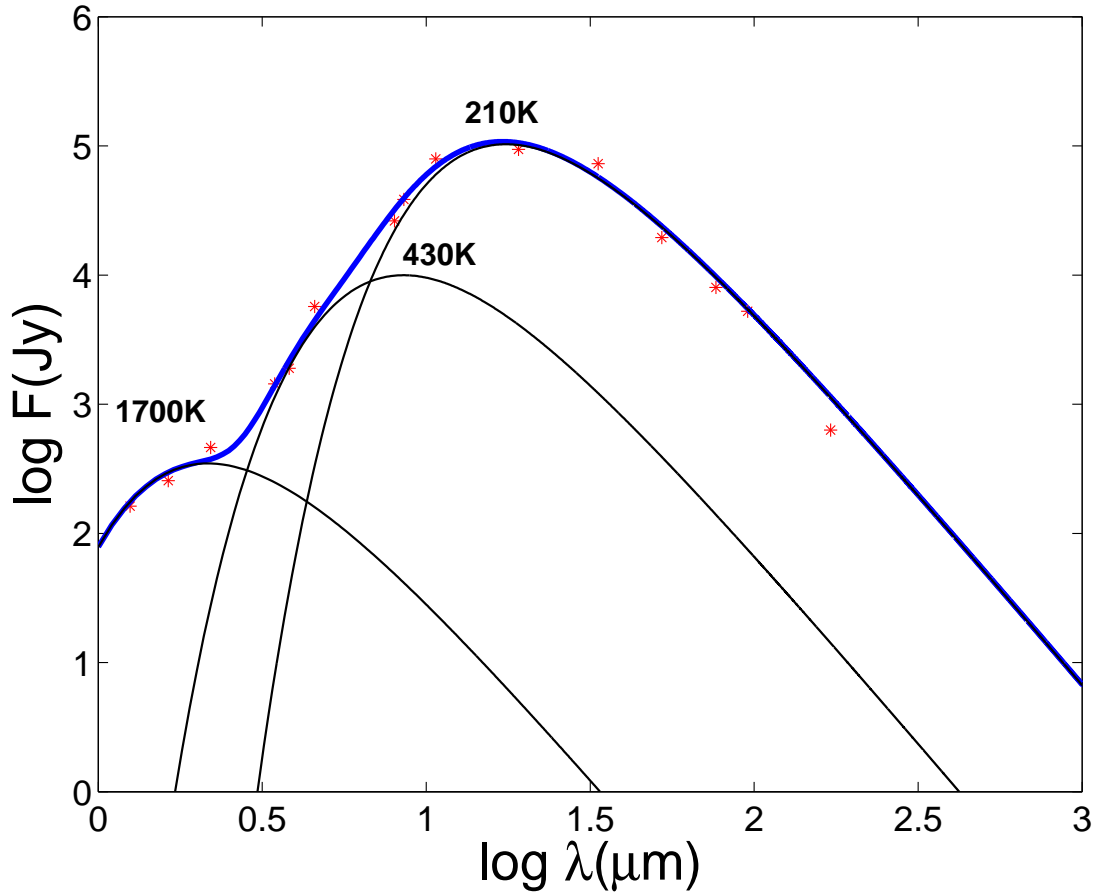}}
\caption{Three temperature modified blackbody fit to the IR. The
points for the \emph{JHKL} wavebands (4 shortest wavebands) are the
average quiescent values from Whitelock et al. (2004), before the
beginning of the event. The rest of the points are taken from
different sources, summarized in Cox et al. (1995). Each modified
blackbody is presented by a thin black line, and the overall
adjusted flux is presented by a thick blue line.} \label{threeTfit}
\end{figure}

We take the average quiescent flux of \emph{JHKL} discussed
in section \ref{sec:models}, add it to the data collected by Cox et
al. (1995), and create a 3-Temperature modified blackbody fit to the entire IR
range
\begin{equation}
F (\lambda) = \rm{b_{210}} \it{\tilde{B}}\rm{(210\K)} + b_{430}
\it{\tilde{B}}\rm{(430\K)} + b_{1700} \it{\tilde{B}}\rm{(1700 \K)}.
\label{F2}
\end{equation}
The coefficients $\rm{b_i}$ were obtained by the method of least
squares adjustment. The results are presented in Fig. \ref{threeTfit}.

\begin{figure}[h!]
\resizebox{0.49\textwidth}{!}{\includegraphics{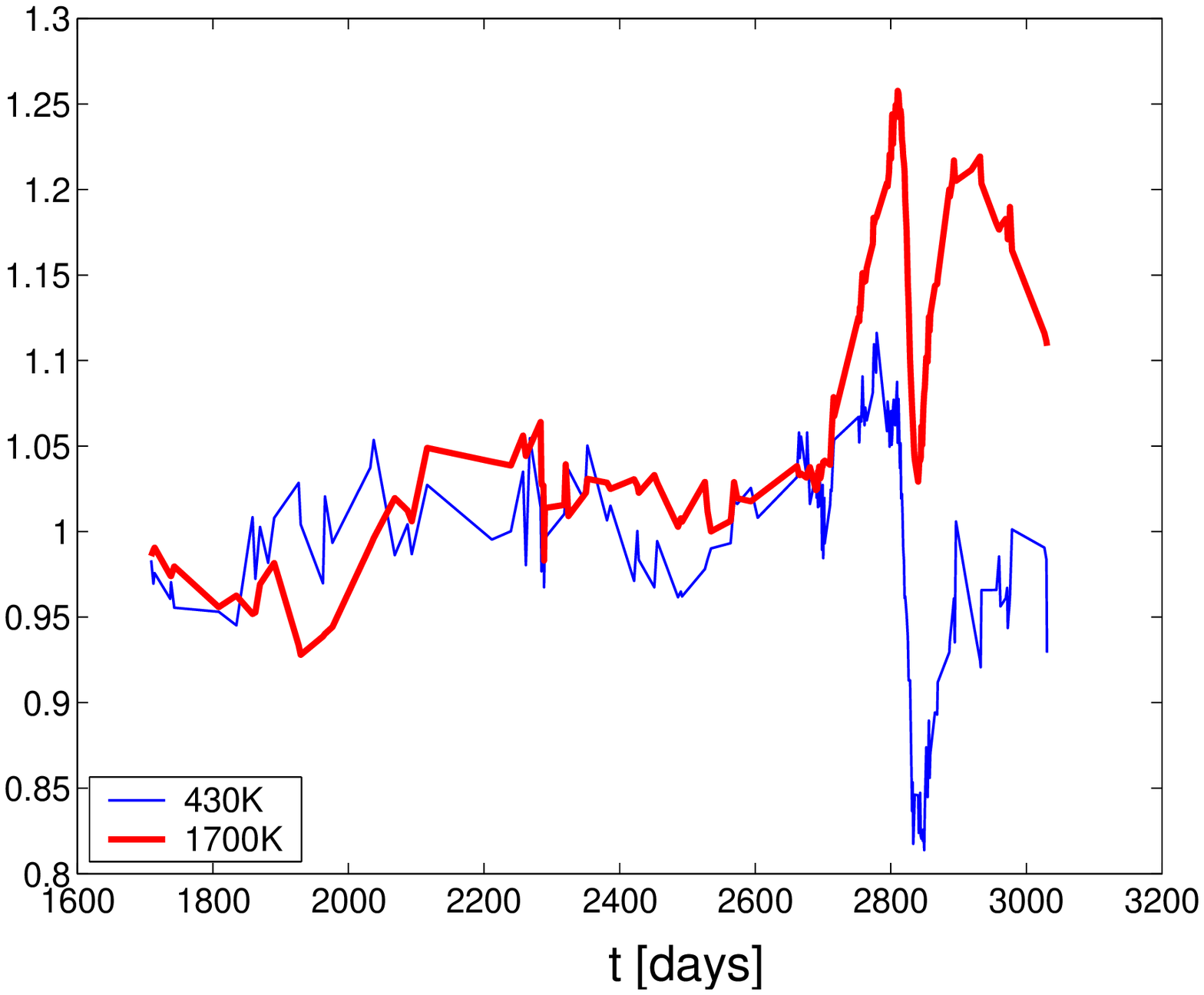}}
\resizebox{0.48\textwidth}{!}{\includegraphics{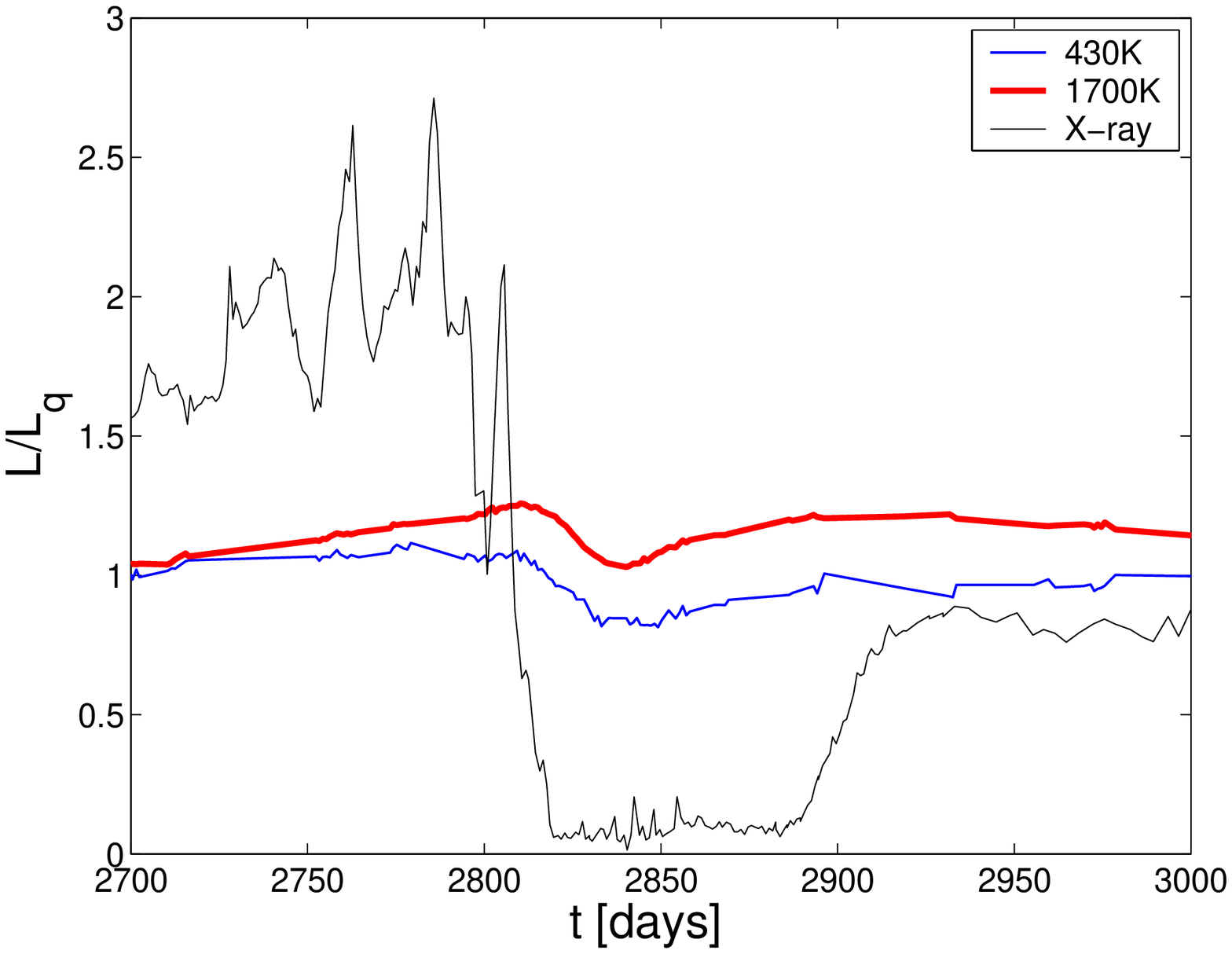}}
\caption{The luminosity of the hot ($1700 \K$; upper red line) and warm
($430 \K$; lower blue line) dust components in units of their
respective average quiescent value $L_q$. The left panel show an
extended period, while the right panel zoom in on the 2003.5 event,
together with the normalized X-Ray flux (from Corcoran 2005). The X-Ray
flat minimum begins at $t=2819.6$.} \label{dustL}
\end{figure}

To better understand the behavior of the \emph{JHKL} emission during
the cycle, and in particular near the event we fit the flux from
Whitelock (2004) only with the two higher temperature components,
and as a function of time
\begin{equation}
F(\lambda,t)=\rm{b_{430}}\it{(t)}\it{\tilde{B}}\rm{(430\K)}+\rm{b_{1700}}\it{(t)}\it{\tilde{B}}\rm{(1700\K)}.
\label{F3}
\end{equation}
We then converted these coefficients to the luminosity of each dust
component by integrating over wavelength the modified blackbody
flux, and multiplying by $4\pi D_{\eta}^2$. The luminosity of the
warm and hot dust components as function of time, and in units of
their respective quiescent average,  is given in Fig. \ref{dustL}.

\section{DUST PROPERTIES}
\label{sec:dustproperties}

Using standard references (e.g., Hildenbrand 1983; Krugel 2003; Whittet
2003) we take the following steps. We assume spherical dust grains with
radius $a = 0.25 \mu m$ and density $\rho_d\sim 1.9 \g\cm^{-3}$
(Hildenbrand 1983). The ratio $Q/a$, where $Q$ is the coefficient of
the modified blackbody emission (equation \ref{tildeB}), is obtained
from figure 8.1 of Krugel (2003), by approximating the wavelengths'
range $1 \mu m \leq \lambda \leq 100 \mu m$
\begin{equation}
\frac{Q}{a} \simeq \frac{7.93}{\lambda}. \label{Q1}
\end{equation}
For our assumed grain radius $a \sim 0.25 \mu m$ (Hildenbrand 1983;
Andersen 2007) we find $\lambda_0 \simeq 2 \mu m$ (see equation
\ref{tildeB}).

We will now consider a spherical dust grain with a radius $a$,
distance $r_d$ from the star, and at equilibrium with the stellar
radiation. The grain is heated mainly by the UV radiation of the
star, which has luminosity $L_\ast$. At the UV band the absorbtion
factor for dust is $Q \sim 1$. The cross section of the spherical
dust grain is taken as $\sigma= \pi a^2$, although it is considered
a bit larger in some models. The dust will emit the radiation
isotropically as the modified blackbody (equation \ref{tildeB}). The
thermal equilibrium is expressed as
\begin{equation}
\sigma \frac{L_\ast}{4 \pi r_d^2}=4 \pi a^2 \cdot \pi c
\int_0^\infty \tilde{B}(T) \lambda^{-2} \,d\lambda. \label{td1}
\end{equation}
where $T$ is the dust temperature. After integrating over the
modified blackbody, we get
\begin{equation}
r_d=26.8 \AU \left(\frac{L_\ast}{9\times10^5
L_{\odot}}\right)^{\frac{1}{2}}\left(\frac{T}{1700
\K}\right)^{-\frac{5}{2}},
\label{td2}
\end{equation}
where we scaled the stellar luminosity with the assumed secondary's
luminosity. The reason is as follows. We assume that most of the
variation in the hot dust luminosity is from hot dust formed in the
shocked primary wind as it collides with the secondary wind. The dense
gas of the primary wind close to the primary, and in the conical shock
region formed by the wind collision absorb most of the primary
radiation (Kashi \& Soker 2007a). Cold dust which is distributed at
larger distances is heated mainly by the primary stellar radiation, and
possibly by the secondary radiation that was not absorbed by the hot
dust. Hence we take $L_\ast=L_1+L_2=5 \times10^6 L_{\odot}$ (Cox et al.
1995; Davidson \& Humphreys 2000) for $T=430\K$ and $T=210\K$. We take
two cases for the hot dust at $T=1700\K$ : $L_\ast=L_2=9\times10^5
L_{\odot}$ (Verner et al. 2005) as a lower limit, and $L_\ast=L_1+L_2$
as an upper limit. The lower limit gives $r_d=26.8 \AU$, as seen from
equation \ref{td2}, and the upper limit gives $r_d=63.1 \AU$. Van
Genderen et al. (1994) found $r_d=63 \AU$ for $T=1700\K$ using regular
blackbody function with totally different parameters which were
acceptable at the time.

We can also evaluate the total mass of the dust. By taking the
luminosity that $N_d(T)$ grains would produce equal to the observed
luminosity obtained by integrating the adjusted flux of each
component (equation \ref{threeTfit}) over wavelength
\begin{equation}
L_d (T)= c \int_0^\infty \rm{b_{i}} \it{\tilde{B}}(T)
\rm{\lambda^{-2}} \,d\lambda , \label{Ld1}
\end{equation}
we can find the number of grains from the equation
\begin{equation}
L_d(T)= N_d(T) 4 \pi a^2 \cdot \pi c \int_0^\infty \tilde{B}(T)
\lambda^{-2} \,d\lambda , \label{md1}
\end{equation}
and then find the mass of the dust from
\begin{equation}
M_d(T)=N_d(T)\frac{4\pi}{3}a^3\rho_d . \label{md2}
\end{equation}
Taking $\rho_d\sim 1.9 \g\cm^{-3}$ (Hildenbrand 1983) for the grain
density, and the following observed quiescent luminosity for each
dust temperature $L_d (1700 \K)=1.1\cdot10^{5}L_\odot$, $L_d (430
\K)=7.7\cdot10^{5}L_\odot$, and $L_d (210
\K)=3.9\cdot10^{6}L_\odot$, we find the approximate dust mass (only
dust, not the inferred total gas mass) $M_d (1700 \K) \simeq 7.6
\times 10^{-9} M_{\odot}$, $M_d(430 \K) \simeq 1.4  \times10^{-5}
M_{\odot}$, and $M_d (210 \K) \simeq 8.4 \times10^{-5} M_{\odot}$.

In an attempt to estimate the mass of dust that resides at a
distance $r_d(T)$ from the binary system we assume that gas flows
from the binary system at the terminal velocity of the primary wind
$v_1 \simeq 500 \rm{km}~\rm{s}^{-1} $. The time it would take for
the gas to arrive at distance $r_d$ from the star is $t_{\rm
flow}(T)=r_d(T)/v_1$.
For the primary mass loss rate we take $\dot{M_1}=3\times10^{-4}
M_{\odot}$, and the conical shock to cover $\sim 1/4$ of the space around the
star (see Kashi \& Soker 2007a).
Assuming the gas to dust ratio is $\sim 100$, the estimated
dust mass from flow consideration is
\begin{equation}
M_d^{\rm flow}(T) \simeq \frac{M_{gas}(T)}{100} \simeq
\frac{1}{400}t_{\rm flow}(T)\dot{M_1}=
7 \times 10^{-7}  
\frac{r_d(T)}{100 \AU} \frac{\dot M_1}{3 \times 10^{-4} M_\odot
\yr^{-1}}  M_\odot \label{mflow}
\end{equation}
In this formula the flow time $t_{\rm flow}(T)$ depends on the distance
of the dust, which in turn depends on the dust temperature.
We summarize our findings in Table 1. 
\begin{table}
\smallskip
\begin{tabular}{||l||r|r|r|r|r||}

\hline \hline

$T(\K)$&$r_d(AU)$&$L_d(L_{\odot})$&$M_d(M_{\odot})$&$t_{\rm flow}(\yr)$&$M_d^{\rm flow}(M_{\odot})$\\
\hline \hline
$1700$ (lower limit)&$26.8$&$1.1\times10^{5}$&$7.6\times10^{-9}$&$0.25$&$ \sim 2\times10^{-7}$\\
$1700$ (upper limit)&$63.1$&$1.1\times10^{5}$&$7.6\times10^{-9}$&$0.60$&$ \sim 5\times10^{-7}$\\
$430$&$1,960$&$7.7\times10^{5}$&$5.3\times10^{-5}$&$18.6$&$\sim 1.5\times10^{-5}$\\
$210$&$11,756$&$3.9\times10^{6}$&$9.6\times10^{-3}$&$111.5$&$\sim 8 \times10^{-5}$ \\
\hline \hline
\end{tabular}
\caption{ The properties of the three dust components. $T(\K)$ is the
dust temperature assumed in our three dust components model. $r_d$ is
the distance of the dust to the star, calculated by assuming that the
hot dust is heated either by the secondary star or both stars, while
the warm and cold dust components are heated by the primary and some of
the secondary radiation (primary and secondary luminosities are given
in the text). $L_d$ is the luminosity of each component as calculated
from the observed IR flux (equation \ref{Ld1}). $M_d$ is the dust mass
calculated from the observed luminosity by equations (\ref{md1}) and
(\ref{md2}). The last two columns examine whether the present primary
wind can supply the dust. Therefore we take the outflow velocity to be
$v_1 \simeq 500 \rm{km}~\rm{s}^{-1} $. $t_{\rm flow}=r_d/v_1$ is the
flow time of the gas from the binary system to the dust location $r_d$.
$M_d^{\rm flow}$ is the dust mass residing near $r_d$ crudely
calculated from the flow time according to equation (\ref{mflow}). The
results for the $210 \K$ and $430 \K$ dust show that the present wind
cannot supply the dust. Those components resides in the Homoncolus and
came from the 19th century eruptions. The $1700 \K$ dust forms close to
the binary system and can be supplied by the present wind.
 } \label{TabelDust}
\end{table}

We emphasize that the last two columns of Table 1 are calculated for
the (wrong) assumption that the dust is supplied by the present wind.
Clearly, as we explain below and is well known (Smith et al. 2003b),
the warm and cold dust cannot be supplied by the present wind.
Therefore, for example, the flow time to the warm dust of $\sim 18.6$
years is not the real flow time. It is the flow time had the dust been
supplied by the present wind. The hot dust can be supplied by the
present wind. Although our "toy model" does not intend to give exact
results to the distances and ages of the dust, but only to give a
general picture, we bring and discuss Table 1 here because it
emphasizes the differences between the hot dust on one hand, and the
warm and cold dust components on the other hand.

{} From Table 1 we learn that the cold dust, which is responsible for
most of the observed IR radiation from $\eta$ Car (Cox et al. 1995),
resides at very large distances from the star, compatible with its
ejection during the Great Eruption that ended $\sim 150$ years ago
(Smith et al. 2003b; Smith \& Gehrz 1998). A small part of this dust
was ejected during the Lesser Eruption $\sim 115$ years ago, which
created circumstellar gas at distances of a few hundred to a few
thousand AU from the star, mainly in the polar directions, but some in
the equatorial plane as well (Smith et al. 2005). Indeed, the present
regular flow from the primary cannot supply the mass required to
explain the emission by the cold dust, as $M_d^{\rm flow} \ll M_d$. As
stated earlier, our analysis of the cold dust is not accurate, but
meant only to present the basic differences between the hot and cold
dust, e.g., that the cold dust cannot be supplied by the present mass
loss from $\eta$ Car. A proper analysis of the cold dust, that includes
also colder dust at $140 \K$, yields even larger dust mass of $\sim 0.1
M_\odot$ (total gas and dust mass of $\sim 10 M_\odot$) in the cold
phase (Smith et al. 2003b). The required cold and very cold dust mass
of $\sim 0.1 M_\odot$ (a gas mass of $\sim 10 M_\odot$) could be easily
supplied by the mass lost during the great eruption (Smith et al.
2003b; Smith 2006; Smith \& Ferland 2007). The total luminosity of all
three components $L_{tot}=4.75\times10^6L_{\odot}$ is smaller than the
assumed luminosity of $\eta$ Car $L_{\eta}=5\times10^6L_{\odot}$ (Cox
et al. 1995). This and our finding that the dust mass can be much
smaller than that ejected during the great eruption, implies that some
radiation escapes without being absorbed by dust. Probably, mainly
along the polar direction (Smith et al. 2003a).

Smith et al. (2003b) have also mapped the dust of $\eta$ Carinae.
They used three different temperatures ($140 \K, 200 \K, 400 \K$),
in their goal to find the total mass and IR luminosity of the
Homunculus. For a component at $T=200\K$ they got luminosity
$L_{200}=1.2\times10^6 L_{\odot}$ and dust mass
$M_{200}=0.015M_{\odot}$, for a component at $T=400\K$ they got
luminosity $L_{400}=5.5\times10^5 L_{\odot}$ and dust mass
$M_{400}=2\times10^{-4}M_{\odot}$. Their total IR luminosity was
$L_{IR}=4.3\times10^6 L_{\odot}$ (Smith et al. 2003b). Although
using some different parameters our results are quite close to their
results.

{ From } Table 1 we also learn  that a large part of the warm dust can
be formed in the outflow from the primary star, and all the hot dust
forms from the primary outflow relatively close to the binary star.
In the next section we propose that the hot dust component and part
of the warm dust component are formed in the conically shaped
post-shock primary wind region. From Table 1 we learn also that if
for some reason the outflow from the binary system changes its
properties, the change in the emission of the hot dust will lag by
$\sim 0.25~{\rm year}=3~{\rm months}$. This time is longer than the
$\sim 10$~weeks duration of the event. This suggests that the
changes in the IR flux during the event is not due to changes in the
outflow that forms the dust. The much slower increase in the IR flux
starting $\sim 150~{\rm days}$ before the event can be accounted for
by changes in the flow properties, as we explain in the next
section. We attribute the rapid changes in the IR fluxes to
obscuration of the secondary star by the mass accreted onto it
during the event (sections 6.1 and  7).

\section{HOT DUST FORMATION}
\label{sec:dust}

Hot dust is present at all times as is inferred from the \emph{J} and
\emph{H} bands. It is probably formed in the primary wind, both the
shocked and unshocked segments. Here we don't consider the dust
formation in the primary wind in quiescence. We only examine the
possibility that the increase in the near infrared luminosity close to
periastron passage is caused by enhanced dust formation in the shocked
primary wind.

\begin{figure}[h!]
\resizebox{0.89\textwidth}{!}{\includegraphics{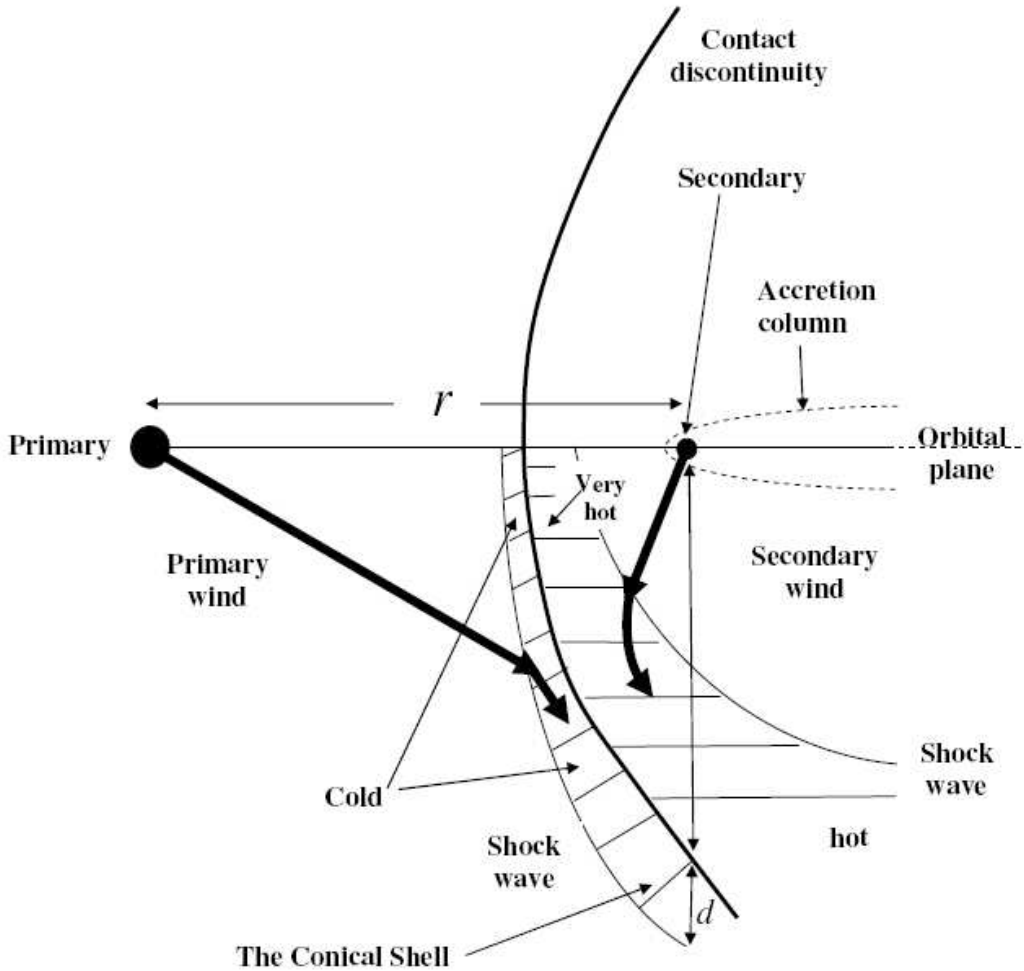}}
\caption{Schematic drawing of the collision region of the two
stellar winds and the definition of several quantities used in the
paper. There is an axial symmetry around the line through the two
stars. The two thick lines represent winds' stream lines. The two
shock waves are drawn only in the lower half. The post-shock regions
of the two winds are hatched. The shocked primary wind region is
referred to as the `conical shell'. The dashed line shows the
accretion column which exists, according to the proposed model, only
for $\sim 70-80~$days during the accretion period which corresponds
to the X-ray minimum and the spectroscopic event.} \label{CWS}
\end{figure}

The two stellar winds collide and go through two respective shock
waves, with a contact discontinuity between the two post-shock wind
regions, as drawn schematically in Fig. \ref{CWS}. The post shock
primary wind cools fast as it outflows (Kashi \& Soker 2007a; Soker
2003; Pittard \& Corcoran 2002) and forms a thin shell in a conical
type shape, which we refer to as the conical shell. We expect dust to
form in this cool high density conical shell. The shocked secondary
wind has a very low density and does not cool. No dust will be formed
in that region. The density in the conical shell depends on several
parameters, e.g., the properties of the secondary winds (Kashi \& Soker
2007a). More influential and relevant to the present study is the
dependence of density on the primary mass loss rate $\dot{M_1}$, on the
orbital separation $r$, on the distance from the primary $r_1$, and on
the magnetic field in the primary wind via its influence on the
post-shock density (Kashi \& Soker 2007a). Typical values of the proton
number density in the conical shell are $n_p \simeq 10^{9}-10^{14}
\cm^{-3}$ (Kashi \& Soker 2007a). It is interesting to note that Smith
et al. (2007) require a density of $n_p \simeq 10^{10} \cm^{-3}$ to
exist in the region where the hot ($\sim 1600 \K$) dust is formed in SN
2006jc.

We examine the ionization structure of the conical shell as a result of
the secondary radiation. Our motivation is to examine whether extra hot
dust can be formed in the conical shell as the system approaches
periastron, in addition to dust expected to be formed in the primary
wind. We neglect the ionizing radiation from the primary because most
of it is assumed to ionize the primary wind close to its origin (Kashi
\& Soker 2007a). We identify three stages of ionization along the
orbit: ($i$) When the orbital separation is large, as is the case for
most of the orbital period, the density in the conical shell is low,
and reaches only $n \sim 10^9-10^{10} \rm{cm^{-3}}$ in the densest
parts. Therefore, the hot secondary ionizing radiation manages to keep
the conical shell totally ionized. In that stage dust can be formed
only at very large distances. Not much hot dust will be formed close to
the secondary. The hot dust might be formed in the primary wind and be
heated by the primary and secondary radiation which penetrate the
conical shell. ($ii$) In a calculation described in Appendix B below,
we find that when the orbital separation decreases to $r \la 14.3 \AU$
at $\sim 170$~days before (after) periastron, the density of the gas in
the conical shell increases such that the cone starts to be optically
thick to the secondary's UV radiation and parts of the cone begin to be
neutral. First at large distances, and then closer and closer to the
stagnation point along the line joining the two stars. ($iii$) As the
secondary approaches periastron, $r \la 3.2 \AU$ at $\sim20$~days
before (after) periastron the ionization front is within the conical
shell in all directions, and the region near the stagnation point
becomes neutral, as shown in Figs. \ref{TSAO} and \ref{cone_ionization}
in Appendix B.

According to the parameters used here, stage ($ii$) starts $\sim
170$~days before periastron passage. We propose that the favorable
conditions for dust formation lead indeed to dust formation close to
the secondary star. This hot dust is the explanation for the rapid
increase in the near-IR starting $\sim 150$~days before the 2003.5
event. Considering the many uncertainties, e.g., the value of the
magnetic field in the primary wind (see Kashi \& Soker 2007a) and the
many fluctuations in the near-IR light curve (Whitelock et al. 2004),
we consider satisfactory that we find the favorable conditions for hot
dust formation, hence the rapid increase in the near-IR flux according
to our model, to start $\sim 170$~days before periastron, while the
observations show it to start $\sim 150~$ days before periastron
passage in the 2003.5 event. The exact time when the conditions become
favorable for dust formation is very sensitive to several of the winds
parameters, and therefore we cannot predict the exact time when the
favorable conditions for dust formation will start in future events.
Indeed, in the 1992.5 and 1998 events there is no rapid increase in the
near-IR flux before the event, but rather a gradual increase that seems
to be part of the secular variation (Whitelock et al. 2004). In the
1981 event there is a rapid rise before periastron, while in the 1987
event the rapid rise occurs earlier on, starting $\sim 300~$days before
the event and finishing after several tens of days. The secular
(gradual) increase in the near-IR flux is observed to occur over the
last 4 decades (Whitelock et al. 2004; Whitelock, P. 2007, private
communication). In addition to the gradual increase there are slow
large flux fluctuations during the orbital period. These fluctuations
and the gradual increase will be treated in a separate paper, where the
secular increase in the near-IR emission will be attributed to the
recovery of \emph{both} the primary (from its high mass loss rate) and
the secondary (from its high mass accretion rate) from the `Great
Eruption'.

The orbital separation $\sim 170$~days before periastron passage
is $r \sim 14.3\AU$. The extra hot dust starts to be formed at a
distance of  $r_2 \sim 13.4\AU$ from the secondary. This is
compatible with the distance of the hot dust from the secondary
$r_2 \sim 25-60 \AU$ as we found in section \ref{sec:dustproperties}
(see Table 1).

The hydrogen and helium visible and infrared lines are no problem in
explaining near-IR emission from dust, since they are not generated in
the same place where the dust resides. The HeI lines are formed in the
acceleration zone of the secondary wind (Kashi and Soker 2007b) while
the dust emitting the near-IR is formed in the outer regions of the
conical shock.

Walsh \& Ageorges (2000) discuss dust features in $\eta$ Car, mainly in
the Homunculus. The IR spectrum of $\eta$ Car is characterized by a
peak around $10 \mu m$ (Mitchell \& Robinson, 1978), indicative of
silicate grains (Whittet 2003). Walsh \& Ageorges (2000) claim that few
percents of the visual and \emph{JHK} radiation is polarized. Polarized
radiation is a strong indication for non-spherical grains. Magnetic
fields make the grains align more or less with their long axes parallel
to each other, causing polarized radiation scattering (Whittet 2003).

The existence of non-spherical silicate grains instead of spherical
amorphous-carbon grains, may slightly change $Q(\lambda)$,  because the
extinction curve for grains from different materials and shape is
different (Krugel 2003). Consequently the curve of the
modified-blackbody function we used (equations \ref{tildeB} and
\ref{Q1}) is changed. These small changes, however, cannot cause a
serious change in our results.

\section{THE ACCRETION PHASE}
\label{sec:accretion}

In total, the accretion process lasts $\sim 10$~weeks (Soker 2005;
Akashi et al.\ 2006). We consider two modes of the accretion
process; we expect the real accretion process to include both. The
discussion below is based on previous papers of the series, in
particular Soker (2005b) and Akashi et al. (2006).

\subsection{General Considerations}
 As was explained in previous papers in the series (Soker 2005b;
Akashi et al. 2006), there are strong arguments supporting an accretion
phase lasting $\sim 10~$weeks near periastron passage. The disk has
time to establish itself because its Keplerian time $t_{dd}$ is much
shorter than the length of the accretion phase. For a secondary mass of
$M_2=30 M_\odot$ the Keplerian (dynamical) time at radius $r_{dd}$ in
the disk is $t_{dd} \simeq 1.4 (r_{dd}/60 R_\odot)^{3/2}~$week. We took
the accretion disk outer radius to be about three times its inner
radius (the secondary radius). The parts of the disk closer to the
secondary surface establish themselves in less than a week.

\begin{figure}[h!]
\resizebox{0.49\textwidth}{!}{\includegraphics{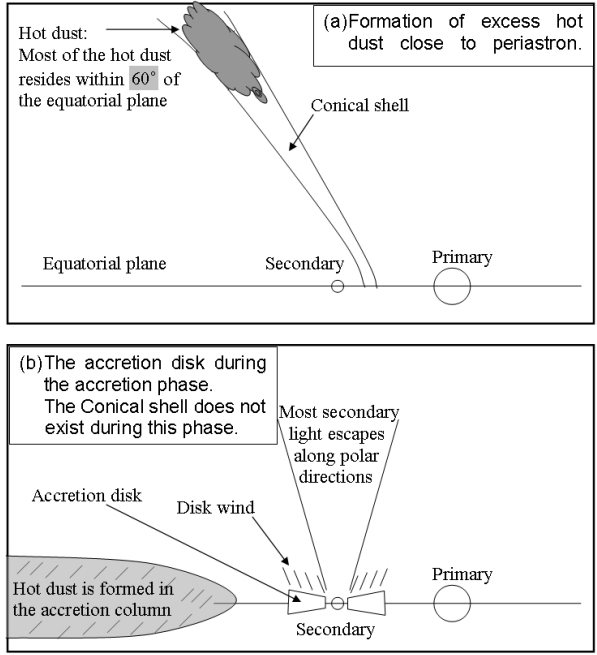}}
\caption{ Schematic geometry of the hot dust and secondary illumination
of the hot dust. Some components are drawn only on one side of the
equatorial plane. Upper panel: When the conical shell, formed by the
winds' collision, exists, then the secondary illuminates it in all
directions. Lower panel: During the accretion phase. The accretion disk
and its wind block the secondary radiation in most directions;
radiation escapes in narrow cones along the polar directions. The
secondary wind does not exist, or exists only along the polar
directions. Hot dust is formed in the accretion column. } \label{newf}
\end{figure}

However, the viscous time scale is longer, and the disk does not
reach a steady state. The viscosity interaction time to spread
material in geometrically thin disks is $t_{vd} \sim 0.1
r_{dd}^2/(\alpha_{dd} C_s H)$ (Frank et al. 1985), where $H \simeq
(C_s/v_K) r_{dd}$ is the vertical size of the disk, $v_K$ is the
Keplerian velocity at $r_{dd}$, $C_s$ is the sound speed, and
$\alpha_{dd} \sim 0.01-1$ is the viscosity parameter. This gives
that the viscous time scale is
\begin{equation}
t_{vd} \simeq t_{dd} \frac{0.1}{ 2 \pi \alpha_{dd}}
\left( \frac{r_{dd}}{H} \right)^2.
\end{equation}
For $H/r_{dd} \simeq 0.3$ and $\alpha_{dd} \simeq 0.1$, we find the viscous time scale
to be somewhat longer than the dynamical (Keplerian) time scale.
The viscous time scale might become as long as a few weeks
at the outer boundary of the accretion disk (although the time is much shorter close
to the secondary surface).
In our model, therefore, a disk is formed, but it does not reach a complete
equilibrium, and does not settle into a fully thin disk.
The disk probably has a wind as part of the evolution toward equilibrium.
The disk and its wind block the radiation in most directions, allowing most of the secondary
radiation to escape only in narrow cones in the polar directions.
This is drawn schematically in Figure \ref{newf}.

For our purpose any dust residing outside the illuminating cones, i.e.,
away from the polar directions, is practically close to the equatorial plane
(even if geometrically speaking it is not). In particular, this holds for dust
that is/was formed in the conical shell.
Therefore, most of the excess hot dust that is being formed in the conical shell
close to periastron passage  will not be heated by the secondary radiation during the
accretion phase.
The same holds for dust that is being formed in the accretion column during the
accretion phase.

The secondary luminosity constitutes $\sim 18\%$ of the total
luminosity in our model (see section \ref{sec:dustproperties}). Because
the secondary is closer to the hot dust and a substantial fraction of
the primary radiation is blocked from heating the excess hot dust (that
was formed in the conical shell) by the primary's own dense wind, the
secondary share in heating the excess hot dust is much larger than its
$\sim 18 \%$ contribution to the bolometric luminosity of the system.
Therefore, during the accretion phase the heating of the hot dust can
potentially be reduced more that $\sim 20 \%$. As we will discuss
below, this is indeed the case.

\subsection{The collapse of the conical shell}

\begin{figure}[h!]
\resizebox{0.49\textwidth}{!}{\includegraphics{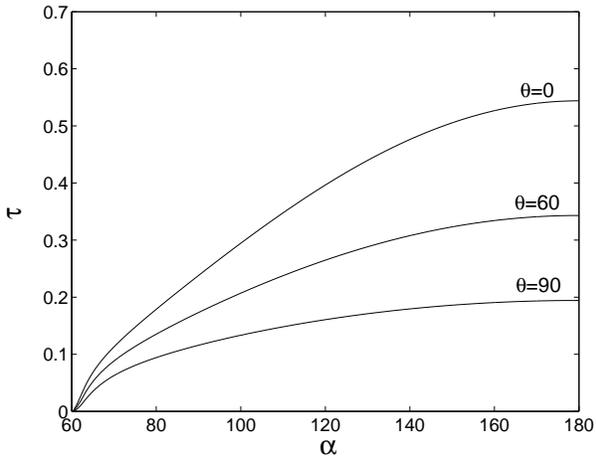}}
\caption{The optical depth of the conical shell as a function of
direction $\alpha$ and for different orbital angles $\theta$:
$\theta=90^\circ$ and $\theta=60^\circ$ corresponding to $\sim 20$ and
$\sim 9$ days before periastron passage, and $\theta=0$ at periastron
passage. $\alpha=180^\circ$ corresponds to the stagnation point. }
\label{tau}
\end{figure}

In the first mode of accretion the conical shell close to the
stagnation region collapses first onto the secondary. The optical depth
of the conical shell at three orbital phases and as function of
direction from the secondary is shown in Fig. \ref{tau}. For the half
space of $\alpha >90^\circ$ close to periastron passage ($\theta=0$) we
take an average value of $\tau_0 \simeq 0.4$. As this region collapses
toward the secondary the optical depth increases as $\tau \ga \tau_0
r_{2s}/r_2$, where $r_{2s} \simeq 0.5 \AU$ and $r_2$ are the distances
of the shell from the secondary before the collapse starts and after,
respectively. The reason the opacity increases faster than $1/r_2$ is
that the density increases as $1/r_2^2$ as the conical shell collapses,
and with it the opacity. We find that $\tau \simeq 1$ at $r_2 \simeq
0.3 \AU \simeq 3 R_2$. Because of the non-negligible angular momentum,
the mass will be accreted mostly from near the equatorial plane (Akashi
et al. 2006).

The stagnation point is at a distance of $\sim 0.5 \AU$ from the
secondary, but regions away from the stagnation point are at larger
distances. Taking the last region to collapse to fall from $\sim 1
\AU$, we find the free fall time to be $\sim 12$~days, which is
about half the duration of the fast decline phase.

\subsection{Accretion from the acceleration zone of the primary wind}

In the second mode we assume that near periastron passage most of
the mass lost by the primary is accreted by the secondary $\dot
M_{\rm acc2} \simeq \dot M_1$. We note that the accretion rate can
be even larger as the primary almost fills its Roche lobe. The
primary spin is not synchronized with the orbital motion, and it is
probably slower, the true primary's Roche lobe is smaller than
calculated for synchronized systems. For an effective temperature of
$T_{\rm eff1}=20,000 \K$ and luminosity of $L_1=4.5 \times 10^{6}
L_\odot$ we find $R_1 \simeq 180 R_\odot$. Assuming that the primary
wind velocity $v_1 =500 \km \s^{-1}$ is the
escape velocity from the primary star, and taking
its mass to be $M_1=120 M_\odot$, we find the same radius $R_1
\simeq 180 R_\odot$. The orbital separation between the two stars
near periastron is in the range $a_p \simeq 250-350 R_\odot$ for an
eccentricity of $0.93-0.9$, respectively. Namely, near periastron
$R_1/a_p \simeq 0.5-0.75$. The secondary is within the acceleration
zone of the primary wind, and the accretion rate can be very high.

Assuming that the mass falls on the secondary at the free fall speed
$v_{ff} = (2 G M_2/r_2)^{1/2}$, and taking for the secondary mass
$M_2 =30 M_\odot$, the density of the accreted matter is
\begin{equation}
\rho_{\rm acc} =  \frac{\dot M_{\rm acc2}}{4 \pi r_2^2 v_{ff}} =
3.06 \times 10^{-13} \left(\frac{\dot M_{\rm acc}}{10^{-4} M_\odot
\yr^{-1}} \right) \left(\frac{r_2}{100 R_\odot} \right)^{-3/2} \g
\cm^{-3}. \label{den2}
\end{equation}
For a temperature of $10^4 \K$ we find the opacity to be $\kappa
\simeq 0.3$.
Therefore, the optical depth is
\begin{equation}
\tau \simeq 3 \left(\frac{\dot M_{\rm acc}}{10^{-4} M_\odot
\yr^{-1}} \right) \left(\frac{r_2}{100 R_\odot} \right)^{-1/2}  .
\label{tau1}
\end{equation}
The angular momentum of this material is large, and most of it will
be accreted from the equatorial plane. We have no way to estimate
the total duration of the high accretion rate phase, and have to
assume that it lasts $\sim 5$~weeks, as the duration of the low IR phase.

Although the conical shell does not exist during the accretion phase,
dust continues to form in the accretion column$-$the dense region
formed behind the accreting secondary (see Fig. \ref{CWS} and previous
papers in the series). The high density and high opacity near
periastron passage and during the initial accretion phase implies that
dust will be formed close to the secondary star. This will be hot dust.
Therefore, there are two competing effects at the beginning of the
accretion phase. On the one hand obscuration of the secondary radiation
in the equatorial plane reduces dust heating. This causes the early
decline in the \emph{L} band. On the other hand dust is formed close to
the secondary star. This dust is hot, and contributes mainly to the
\emph{J} band. Initially this effect dominates over the obscuration of
radiation, and therefore the flux in \emph{J} starts to decline later.
The delay between the warm and hot dust is seen also in Fig.
\ref{dustL}, where the X-ray light curve is also presented. The
obscuration becomes the dominant effect when the accretion phase is
well developed, when the X-ray drops as well.

The accretion phase continues after the secondary has left the
acceleration zone of the primary wind, and lasts for a total of $\sim
10~$weeks. The accreted gas at the second half of the accretion phase
has angular momentum that is not enough to form an accretion disk.
However, it is high enough to channel accretion to the equatorial
plane; we continue to refer to this equatorial accretion as `disk'. The
accretion of gas mostly from the vicinity of the equatorial plane
during the entire accretion phase results in a collimated polar outflow
(wind) that after being shocked emits in the hard X-ray, and accounts
for the residual X-ray emission during the event (Akashi et al. 2006
section 4.3.1). The speed of this polar outflow is similar to the
secondary wind speed, hence will lead to similar X-ray temperature, as
found by Hamaguchi et al. (2007; for that the criticism of Hamaguchi et
al. of the accretion model is not valid). We emphasize that we are
referring here to a collimated flow blown by the secondary star. The
primary star continues to blow its much more massive wind, which is
spherical during the event (Smith et al. 2003a).

This polar outflow will make the optical depth in the polar directions
much lower than in the equatorial plane. Hence, a large fraction of the
secondary radiation will be emitted along the polar direction during
the accretion phase. Because all the excess hot dust (that is formed in
the conical shell close to periastron passage) resides within $\la 60
^\circ$ from the equatorial plane, and a large fraction of the hot and
warm dust that is formed in the unshocked primary wind resides away
from the polar directions as well [Smith et al. 2000; cooler dust at $T
<200 \K$ resides in the polar lobes and larger distances (Smith et al.
2003b; Gomez et al. 2006)], we expect the dust emission to decrease as
the disk blocks the secondary radiation; this is the subject of section
\ref{sec:model}. As the spectroscopic event proceeds, the secondary's
distance from the primary increases, and the accretion rate decreases.
The polar opening angle is expected to increase, and more of the
secondary radiation will reach regions closer and closer to the
equatorial plane, where hot dust resides.

\section{DISCUSSION AND SUMMARY: A MODEL FOR THE NEAR-IR LIGHTCURVE}
\label{sec:model}
The main goal of this paper is to explain the fast decrease and then
increase in the near-IR emission that occurs during the spectroscopic
event (Whitelock et al. 2004). For that, we now incorporate all the
ingredients discussed in previous sections, and try to explain the
near-IR light curve near periastron passage.

In section 2 we argued that free-free emission cannot account for the
near-IR behavior. Basically, the free-free emission cannot supply the
required luminosity. We also showed that absorption of the near-IR
emission cannot account for the decline. In sections
\ref{sec:fit_lightcurve} and \ref{sec:dustproperties} we explored the
dust properties, by building a `toy model' where the dust is composed
of three-temperature components. We found there that the hot (1700~K)
dust can be supplied directly by the present mass loss rate of the
primary star. The warm and cold dust, on the other hand, must come
mainly from the eruptions of the 19th century and even from earlier
times (Smith et al. 2003b; Gomez et al. 2006).

In section \ref{sec:dust} we argued that as the two stars approach
periastron conditions become favorable for dust formation in the
conical shell$-$the post shock region of the primary wind (see Fig.
\ref{CWS}); This is presented in Figs. \ref{TSAO} and
\ref{cone_ionization}). We attribute the rapid increase in the near IR
flux starting about five months before the 2003.5 event to enhanced
dust formation in the conical shell. The conditions are very sensitive
to several of the wind parameters of both stars. Indeed, in some past
events there was no rapid rise, while in others the rapid rise occurred
at different orbital phases before periastron (see section
\ref{sec:dust}). We therefore cannot predict the exact time for the
favorable dust formation conditions, hence the rapid increase in the
near-IR flux, to occur in future events. In addition to this
sensitivity there is the gradual change in the IR flux observed for
almost 40 years now, which we attribute in a future paper to the
recovery of \emph{both} the primary and secondary from the Great
Eruption of the 19th century.

We turn to describe the decline during the spectroscopic event. In
previous papers in the series the absorption of the secondary ionizing
radiation was considered (Soker 2007; Kashi \& Soker 2007a). Here we
are also interested in the near UV and visible bands which are
responsible for heating the dust. Our findings in section
\ref{sec:accretion} suggest that the mass accretion rate near
periastron passage is such that the secondary radiation capable of
heating the dust is almost completely blocked in directions away from
the polar directions. Most of the radiation will escape in narrow cones
in the polar directions, and some will be converted to longer
wavelengths by the optically thick accreted gas. These longer
wavelengths are less efficient in heating the dust. We expect some of
the hot dust to be formed in the conical shock (section
\ref{sec:dust}), hence to reside within $\sim 60^\circ$ of the
equatorial plane. The polar radiation will not heat this dust, and the
dust will cool down and the emission in the \emph{JHKL} IR bands will
drop. Dust at very large distances will be heated by the primary's
radiation and secondary's polar radiation, so we expect no decline in
the far IR. (e.g., in $\sim 30 \mu m$).

Most of the excess hot dust, i.e., the dust that is formed within
several months from periastron passage, is heated by the secondary
star, that accounts for $\sim 18 \%$ of the bolometric luminosity.
Therefore, the blockage of the radiation of the secondary will erase
almost completely the increase in the near IR radiation prior to the
decline. Hot dust that is formed close to the secondary will partially
compensate for this decline. More than that, the blocked radiation from
the secondary (during the accretion phase) does not heat the dust that
is formed in the unshocked primary wind. Therefore, the total decline
can bring the radiation to be below its quiescent level.

The high density and high opacity near periastron passage and during
the initial accretion phase implies that dust will be formed close to
the secondary star. This will be hot dust. Therefore, there are two
competing effects at the beginning of the accretion phase. On the one
hand obscuration of the radiation of the secondary in the equatorial
plane reduces dust heating. This causes the early decline in the
\emph{L} band. On the other hand dust is formed close to the secondary
star. This dust is hot, and contributes mainly to the \emph{J} band.
Initially this effect dominates over the obscuration of radiation, and
therefore the flux in \emph{J} starts to decline later. The delay
between the warm and hot dust is seen also in Fig. \ref{dustL}, where
the X-ray light curve is also presented. The obscuration becomes the
dominant effect when the accretion phase is well developed, at that
stage the X-rays sharply drop as well.

We attribute the sharp decline in the near-IR emission near periastron
passage (the V-shaped region in the light curve) to the high mass
accretion phase occurring at the beginning of the accretion phase
(lasting $\sim 5$~weeks). The Bondi-Hoyle accretion rate of the
secondary from the primary wind outside its acceleration zone is
$M_{\rm acc} \simeq 10^{-6} M_\odot$ (Akashi et al. 2006), and the
optical depth is very low. Therefore, in the second part of the
accretion phase (lasting $\sim 5$ weeks,) when the orbital separation
increases, the radiation of the secondary manages to heat more and more
dust closer to the equatorial plane, mainly hot dust, and the near-IR
emission recovers.

The decrease in the heating radiation from the secondary toward low
latitudes starts when matter starts to be accreted on the secondary.
This occurs a few weeks before the event. This explains the early small
decrease in the \emph{L} band. At the same time, more dust is formed
close to the secondary star. This is a hot dust, that more than
compensate which more than compensates for the decrease in the
radiation, hence the flux in the \emph{J} band continues to increase.
Eventually, as the accretion phase is developed, there is a rapid
decline in all bands.

We suspect that after the high mass accretion rate phase starts, it is
mainly the nearby dust that formed in the accretion column that is
being heated. This is a relatively hot dust. The secondary does not
heat dust at large distances in the equatorial plane, which emits much
more in the \emph{L} band. Therefore, \emph{J} recovers relatively more
than \emph{L} (in absolute flux \emph{L} changes more).

At all times the primary continues almost unaffected its role in
heating the dust, therefore the decline in the IR is only $\sim
10-24 \%$.

We are grateful to Patricia whitelock, Freddy Marang and Francois van
Wyk for sending us their data plotted in Fig. \ref{whitelock}. We thank
Nathan Smith for very helpful comments. This research was supported by
a grant from the Asher Space Research Institute at the Technion.

\appendix
\section{Appendix: Free-Free Absorption Calculations:}

\begin{figure}[h!]
\resizebox{0.49\textwidth}{!}{\includegraphics{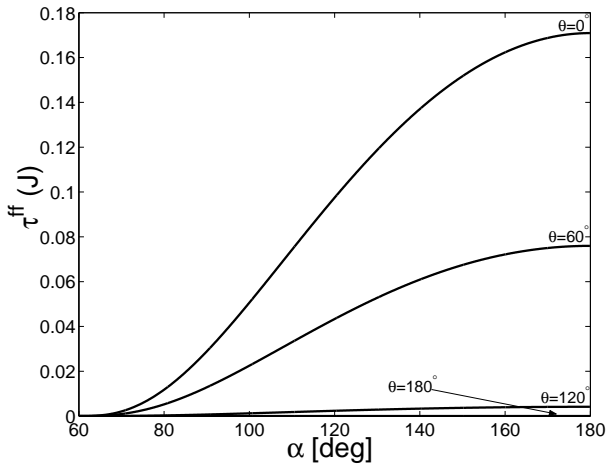}}
\caption{ The free-free optical depth in the \emph{J} waveband as a
function of $\alpha$ and $\theta$ as given by equation \ref{tau_ff_J}.
$\alpha$ is the angle of a ray through the conical shell measured from
the secondary ($\alpha=180^\circ$ toward the primary; see Fig.
\ref{cone_ionization} below), and $\theta$ is the orbital angle
($\theta=180^\circ$ at apastron and $\theta=0^\circ$ at periastron).}
\label{tau_ff_J_fig}
\end{figure}

We now show that the dense post-shock primary wind material (the
conical shell) cannot absorb enough of the \emph{J}-band emission (for
more detail on the absorption by the conical shell see Kashi \& Soker
2007a). The free-free absorption coefficient at frequency $\nu$ is
given by equation (5.18) of Rybicki \& Lightmann (2004)
\begin{equation}
\alpha_{\nu}^{ff}=3.7 \times 10^{8} T^{-\frac{1}{2}} Z^2 n_e n_i
\nu^{-3} \left[1-\exp\left(-\frac{h\nu}{KT}\right)\right]g_{ff}
\cm^{-1} , \label{alpha_ff_RL}
\end{equation}
where, as discussed in the section 2.2, $T=10^4 \K$ and
$g_{ff} \simeq 1.2$. We calculate the free-free optical depth for
the \emph{J}-band as a function of $\alpha$ (the angle between a
direction opposite to the direction of the primary to a point on the
contact discontinuity measured from the secondary; see Fig.
\ref{cone_ionization} below) and $\theta$ (the orbital angle; see
Fig. \ref{TSAO} below). Defining $d$ as the post shock primary wind
depth (see Kashi \& Soker 2007a, and Fig. \ref{cone_ionization}
below) we find that
\begin{equation}
\tau_{\nu}^{ff}=2.2 \times 10^{-37} n_e(\alpha,\theta)
n_i(\alpha,\theta) d(\alpha,\theta), \label{tau_ff_J}
\end{equation}
where $n_e$, $n_i$ and $d$ are in c.g.s. units. Figure
\ref{tau_ff_J_fig} shows The free-free optical depth in the \emph{J}
waveband as a function of $\alpha$ and $\theta$. The maximum value of
$\tau_{\nu}^{ff}$ is obtained for a direction through the primary
($\alpha=180^{\circ}$) at periastron ($\theta=0^{\circ}$). For the
point on the conical shell along this direction the electron and ion
densities are $n_e \sim 1.9 \times 10^{12} \cm^{-3}$, and
$n_i\sim1.6\times10^{12} \cm^{-3}$, respectively, and the geometrical
width of the conical shell along this direction is $d\sim2.5\times
10^{11} \cm$. The maximum value obtained is therefore
$\tau_{\nu}^{ff}\sim 0.17$. We note that this is an hypothetical
maximum, because according to our model the colliding winds cone does
not exist close to periastron passage. We conclude that the dense
post-shock primary wind (the conical shell) is optically too thin to
absorb enough of the \emph{J}-band emission as seen in observations.

\section{Appendix: Ionization and Dust Formation in the Colliding Winds Cone:}

\begin{figure}[h!]
\resizebox{0.89\textwidth}{!}{\includegraphics{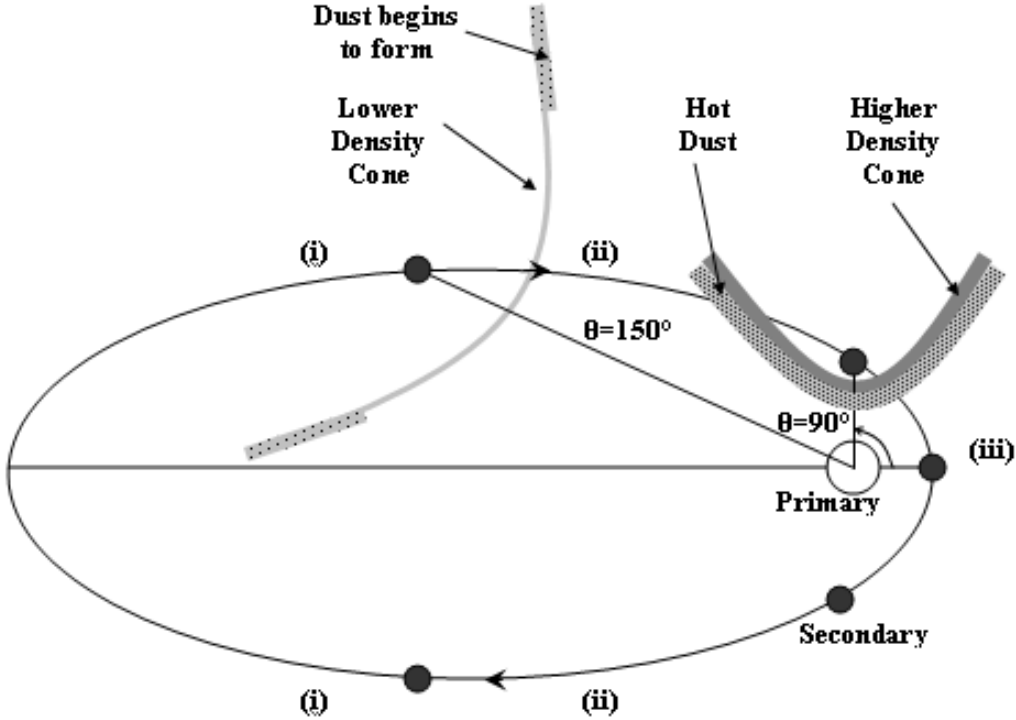}}
\caption{ The three stages of the conical shell ionization plotted
along the secondary's orbit together with the orbital phases of
transitions between stages. Darker colors means denser wind.
Stage($i$): When the orbital separation is large, the hot secondary
ionizing radiation manages to keep the conical shell (the post-shock
primary wind zone) totally ionized. In that stage dust can be formed
only at very large distances. Stage ($ii$): When the orbital separation
decreases to $r \la 14.3 \AU$ at $\sim 170$~days ($\theta=150^{\circ}$)
before (after) periastron, the density of the gas in the conical shell
increases, the cone becomes optically thick to the secondary's UV
radiation and parts of the cone begin to be neutral. First at large
distances, and then closer and closer to the stagnation point along the
line joining the two stars. Stage ($iii$): As the secondary approaches
periastron, $r \la 3.2 \AU$ at $\sim20$~days ($\theta=90^{\circ}$)
before (after) periastron the ionization front is within the conical
shell in all directions, and the region near the stagnation point
becomes neutral. During most of this stage the conical shell does not
exist according to our model. Instead, an accretion flow onto the
secondary takes over. The exact times (and locations) of the
transitions between the stages depend on poorly determined winds'
parameters. }

\label{TSAO}
\end{figure}
\begin{figure}[h!]
\vskip -1. cm
\resizebox{0.33\textwidth}{!}{\includegraphics{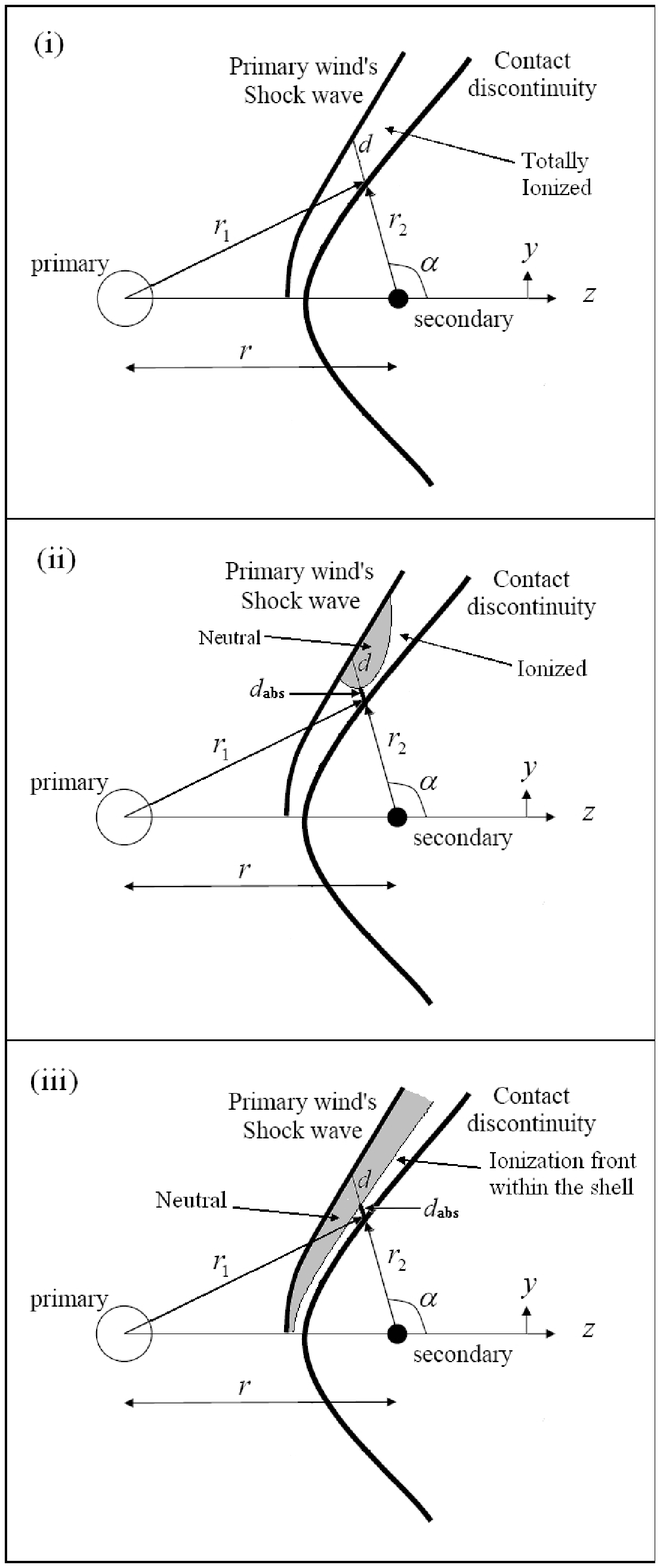}} \vskip
-0.2 cm \caption{Schematic drawing of the plane between the two
stars, at the three stages of the conical shell ionization. $d$ is
the thickness of the shocked primary wind zone (conical shell) from
the contact discontinuity to the primary shockwave in $\alpha$
direction. $d_{\rm abs}$ is the distance along the line from the
contact discontinuity through the shocked primary wind where the
wind is optically thin to the secondary's radiation, in the same
direction. Upper panel $-$ Stage($i$): Until $\sim170$~days before
(after) periastron $d_{\rm abs}>d$ for all values of $\alpha$. The
secondary ionizing radiation manages to keep the conical shell
totally ionized.  Middle panel $-$ Stage($ii$): From $\sim170$~days
until $\sim20$~days before (after) periastron $d_{\rm abs}>d$ for
small values of $\alpha$ and $d_{\rm abs}<d$ for large values of
$\alpha$, and parts of the cone begin to be neutral. Lower panel $-$
Stage($iii$): From $\sim20$~days before periastron until
$\sim20$~days after periastron $d_{\rm abs}<d$ for every $\alpha$
and the ionization front is within the conical shell in all
directions, and the region near the stagnation point becomes
neutral. The last stage exists only for a short period of time, or
might not exist at all, because of the collapse of the conical shell
onto the secondary $\sim20$~days before periastron, and the
formation of an accretion flow (Akashi et al. 2006).}
\label{cone_ionization}
\end{figure}

\begin{figure}[h!]
\resizebox{0.89\textwidth}{!}{\includegraphics{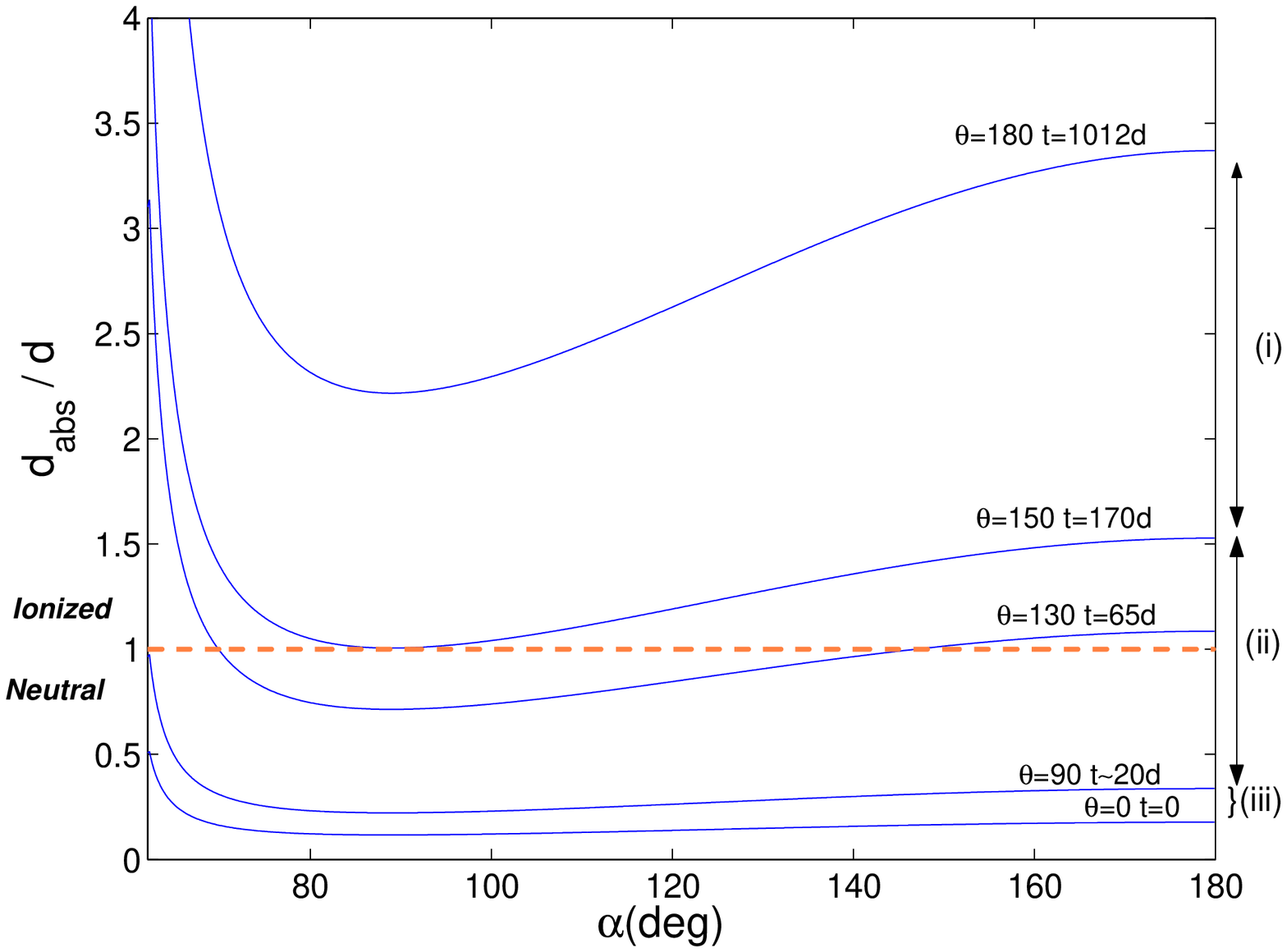}}
\caption{$d_{\rm abs}(\alpha)$ is the distance along the line at angle
$\alpha$ from the contact discontinuity through the shocked primary
wind (conical shell) required to absorb the ionizing radiation of the
secondary star. In our model the asymptotic angle of the conical shell
is at $\alpha=60^\circ$ (left side of the figure), where
$\alpha=180^\circ$ is the angle toward the primary and $\alpha=0^\circ$
in a direction opposite to the primary. $d(\alpha)$ is the total length
through the conical shell along the same direction (see Fig.
\ref{cone_ionization}). If $d_{\rm abs}(\alpha)/d(\alpha) > 1$ then the
conical shell along the direction $\alpha$ is completely ionized, while
if $d_{\rm abs}(\alpha)/d(\alpha) < 1$ part of the conical shell is
neutral. Neutral regions are more favorable for dust formation.
$\theta$ is the orbital angle (see Fig. \ref{TSAO}), with $\theta=0$ at
periastron ($t=0$), and $\theta =180^\circ$ at apastron
($t=1012~$days). The three stages discussed in the text, and the times
(in days) before periastron passage of transitions between the stages
are marked on the figure. } \label{dabs}
\end{figure}

In finding the ionization front location we use our colliding winds
model (Kashi \& Soker 2007a) and assume that dust forms in the
post-shock primary wind region. We define the direction angle $\alpha$
as the angle from a direction opposite to the primary direction to a
point on the contact discontinuity as measured form the secondary (see
Fig. \ref{cone_ionization}). Segments of the conical shell start to be
neutral when the total recombination rate per steradian along a
direction $\alpha$ measured from the secondary becomes equal or larger
than the rate of ionizing photons ($h \nu \geq 13.6 \rm{eV}$) emitted
by the secondary star per steradian, $\dot{N}_{i2}$. For the secondary
of $\eta$ Car $\dot{N}_{i2}=\dot \Phi_{2}/{4\pi} = 2\times10^{48}
\rm{s^{-1} sr^{-1}}$, as derived by Kashi \& Soker (2007a) based on
Schaerer \& de Koter (1997). We neglect recombination in the low
density secondary wind. Thus the distance along the line from the
contact discontinuity through the shocked primary wind where the wind
is optically thin to the secondary's radiation, as can be seen in Fig.
\ref{dabs},  is
\begin{equation}
d_{\rm abs}=\frac{\dot{N}_{i2}}{\alpha_B n_e n_H r_2^2}
\end{equation}
where $\alpha_B$ is the recombination coefficient, $r_2$ is the
distance to the secondary, $n_e$ is the electron number density and
$n_H$ is the proton number density.

We recall that the ionizing radiation from the primary is neglected
here because most of it is absorbed in the dense gas of the primary
wind close to the primary, and in the dense gas in the conical shock
region. Despite being five times as luminous as the secondary, the
primary is much cooler and its ionizing photon flux ($h \nu > 13.6
\eV$) is small to begin with. However, at longer wavelengths in the UV
and in the visible band the primary radiation cannot be neglected, and
it is considered as a heating source of the dust together with the
secondary radiation (see discussion following equation \ref{td2}).

The value of $d_{\rm abs} (\alpha)$ is then compared to the
thickness $d (\alpha)$ of the wind in the same direction $\alpha$.
In stage ($i$) $d_{\rm abs}>d$ for every $\alpha$ and the shocked
primary wind (the conical shell) is totally ionized. In stage ($ii$)
$d_{\rm abs}>d$ for small values of $\alpha$ and $d_{\rm abs}<d$ for
large values of $\alpha$, and parts of the cone begin to be neutral.
In stage ($iii$) $d_{\rm abs}<d$ for every $\alpha$ and the
ionization front is within the conical shell in all directions.
These three stages are drawn schematically in Figs. \ref{TSAO} and
\ref{cone_ionization}.

When the material in the conical shell is neutral, the conditions
are favorable for dust to be formed. The gas in the conical shell is
initially hot $T_{gas}=10^4 \K$. As it expands and cools, hot dust
forms at temperatures $T \la 1700 \K$ (Van Genderen et al. 1994).
The cooling time depends on density. We use figure 11 of Woitke et
al. (1996) to estimate the cooling time from $T=10^4 \K$ and $n
\simeq 10^{12}-10^{14} \cm^{-3}$ to $T=1700 \K$ in an adiabatic
process and find it to be approximately $5-10$~days.

For most of the orbital period the wind is ionized and no dust is
being formed in the conical shell close to the secondary. The
roughly constant IR emission is emitted by dust which was formed in
previous cycles, mostly cooler dust, and the hotter dust that is
formed in the primary wind and heated by the secondary and primary
radiation. The rapid increase in the IR emission $\sim 150 \days$
before the 2003.5 event (see section \ref{sec:model}) can be explained by
the process described in stage ($ii$) - more and more material in
the conical shell becomes neutral, favoring the formation of dust
closer to the secondary. The occurrences of the event are explained
in section \ref{sec:model}.

We stress again that the numbers can be somewhat different for other parameters,
like $e$, stellar masses and magnetic field in the primary wind.

\end{document}